\documentclass[twocolumn, prl, amssymb, superscriptaddress, aps, showpacs,preprintnumbers,
amsmath,floatfix]{revtex4}

\usepackage{epstopdf}
\usepackage{graphics}
\usepackage{graphicx}
\usepackage{dcolumn}
\usepackage{bm}
\usepackage{longtable}
\usepackage{epsfig}
\usepackage{times}
\usepackage{url}
\usepackage{color}
\usepackage{subfigure}

\begin{document}

\title{Collective magnetic splitting in single-photon superradiance}

\author{Xiangjin \surname{Kong}}
\email{xjkong@mpi-hd.mpg.de}
\affiliation{Max-Planck-Institut f\"ur Kernphysik, Saupfercheckweg 1, D-69117 Heidelberg, Germany}
\affiliation{Department of Physics, National University of Defense Technology, Changsha 410073, People's Republic of China}

\author{Adriana \surname{P\'alffy}}
\email{palffy@mpi-hd.mpg.de}
\affiliation{Max-Planck-Institut f\"ur Kernphysik, Saupfercheckweg 1, D-69117 Heidelberg, Germany}

\date{\today}

\begin{abstract}

In an ensemble of identical atoms, cooperative effects like sub- or superradiance may alter the  decay rates and the energy of  specific transitions may be shifted from the single-atom value by the so-called collective Lamb shift.  
While such effects in ensembles of two-level systems are by now well understood, realistic multi-level systems are more difficult to handle. In this work we show that in a system of atoms or nuclei under the action of an external magnetic field, 
the  collective contribution to the level shifts can amount to seizable deviations from the single-atom Zeeman or magnetic hyperfine splitting picture. We develop a formalism to describe single-photon superradiance in multi-level systems in the small sample limit and quantify the parameter regime for which the collective Lamb shift leads to measurable deviations in the magnetic-field-induced splitting. In particular, we show that this effect should be observable in the nuclear magnetic hyperfine splitting in M\"ossbauer nuclei embedded in thin-film x-ray cavities.

\end{abstract}

\maketitle
In 1947, a famous experiment by Lamb and Retherford \cite{Lamb1947} confirmed that the $2s_{1/2}$ and $2p_{1/2}$ levels in hydrogen are not  degenerate, leading to increased efforts in the theoretical understanding of radiation  quantization \cite{Bethe1947,Weisskopf1949} and to the development of quantum electrodynamics (QED) \cite{QEDbook}. Today, it is known that this small shift has to do with emission and reabsorption of virtual photons within the atom, mainly with self-energy and vacuum-polarization corrections. Interestingly, it could be shown that an additional contribution arises if many identical atoms are interacting collectively with resonant photons, and virtual photons are exchanged between different atoms \cite{Friedberg1973,Svidzinsky2009,Scully2009.superradiance,rohlsberger2010collective,keaveney2012cooperative,meir2014cooperative}. This additional contribution has been  termed in analogy {\it collective Lamb shift}, although the exact underlying processes share with the single-atom Lamb shift mechanism only the virtual character of the photons 
being exchanged. 

The collective Lamb shift has been investigated theoretically for ensembles of two-level systems in the small and large sample limits \cite{Friedberg1973,Svidzinsky2010.PRA,svidzinsky2008dynamical,scully2006directed,Manassah2012,Roof2016}. Collective scattering of a weak-intensity laser off a cold ensemble of rubidium atoms with Zeeman splitting has been recently investigated both theoretically  and experimentally \cite{jennewein2016coherent,pellegrino2014observation,jenkins2016collective,lee2016stochastic}, with results that partly contradict predictions of the standard cooperative Lamb shift theory. In the context of ensembles of atoms in magnetic fields, a legitimate question is to which extend can the Zeeman splitting be considered independently from 
the cooperative effects and in particular the collective Lamb shift. In this Letter, we show that in ensembles of atoms or nuclei with magnetic-field induced level multiplets, the collective magnetic  splitting  can show significant deviations of the Zeeman splitting compared to the single-atom behaviour and cannot be justified by mere two-level system collective Lamb shift contributions for each transition independently, as done for instance in Ref.~\cite{jennewein2016coherent}. We investigate theoretically the collective effects for the case of single-photon superradiance in ensembles of atoms or M\"ossbauer nuclei in the small sample limit. Based on our results, we classify the regimes under which the collective magnetic  splitting is relevant and show that this effect should be observable in planar x-ray cavities with an embedded nuclear layer  \cite{rohlsberger2010collective} under experimental parameters available today. Our findings give new and unexpected insights on the collective behaviour of optical and x-ray systems in magnetic fields.

The single-photon cooperative emission from a cloud of $N$ two-level atoms has been subject of sustained interest in the last decade \cite{Scully2006.PRL,Svidzinsky2008.PRA,Zhang2012}. It was shown that in consistence with Dicke's results on superradiance back in 1954 \cite{Dicke1954}, the decay of the excited system is for certain sample geometries proportional with the number of atoms in the sample $\Gamma=N\gamma$, where $\gamma$ is the spontaneous decay rate of a single atom   \cite{Temnov.PRL95,Kaiser.PRL101,Scully2009.superradiance}.  The frequency is shifted by the collective Lamb shift $\mathcal{L}$ compared to the case of a single atom. We now assume that a magnetic field introduces Zeeman splitting of the two-level system. For simplicity, we consider the case when both the ground state $g$ and the excited state $e$ are split in two sublevels only. Using specific polarization directions  one can envisage the situation that only two hyperfine transitions, denoted in the following by 1 and 2, can be driven by the resonant field. The $N_i$ atoms in the ground state 
$i$ with $i=1,2$ and $N_1+N_2=N$ are located at positions  $\vec{r}_j\,\,(j=1,...,N)$. Although the results are derived in the following for a generic ensemble of atoms, this case is in particular also appropriate for a sample of M\"ossbauer nuclei \cite{adams2013,Vagizov2014}, the system that allowed the experimental observation of the collective Lamb shift in single-photon superradiance \cite{rohlsberger2010collective}. 
 The interaction between  atoms/nuclei and photons is described by the Hamiltonian in the interaction picture (with $\hbar=1$)
\begin{align}
H_{int}=&\sum_{\vec{k}}\sum_{j=1}^{N_1}g_{\vec{k},1}(\hat{\sigma}_{1}^{j}e^{-i\omega_{1}t}+\hat{\sigma}_{1}^{j+}e^{i\omega_{1}t}) \nonumber \\
&\times 
(\hat{a}_{\vec{k}}^{\dagger}e^{i\nu_kt-i\vec{k}\cdot \vec{r}_j}+\hat{a}_{k}e^{-i\nu_kt+i\vec{k}\cdot \vec{r}_j}) \nonumber \\
&+\sum_{\vec{k}}\sum_{j=N_1+1}^{N}g_{\vec{k},2}
(\hat{\sigma}_{2}^{j}e^{-i\omega_{2}t}+\hat{\sigma}_{2}^{j+}e^{i\omega_{2}t}) \nonumber \\
&\times
(\hat{a}_{\vec{k}}^{\dagger}e^{i\nu_kt-i\vec{k}\cdot \vec{r}_j}+\hat{a}_{\vec{k}}e^{-i\nu_kt+i\vec{k}\cdot \vec{r}_j})\, ,
\label{H}
\end{align}
where $\hat{\sigma}_{1(2)}^{j}$ is the lowering operator of transition 1(2) for atom $j$, $\hat{a}_{\vec{k}}$ ($\hat{a}^\dagger_{\vec{k}}$)  is the photon annihilation (creation) operator, and $g_{\vec{k},1(2)}$ is the atom-photon coupling constant for the  mode $\vec{k}$. For simplicity, we assume $g_{\vec{k},1}=g_{\vec{k},2}=g_k$. The frequencies $\omega_1$ and $\omega_2$ characterize the two transitions between the single-atom hyperfine-split levels, and $\nu_k$ is the field mode frequency. The rotating wave approximation is not applicable in this case \cite{Friedberg2008,Svidzinsky2008,Friedberg2008Reply}. 
We look for a solution of the Schr\"{o}dinger equation for the atoms and the field as a superposition of Fock states \cite{Svidzinsky2010.PRA}
\begin{align}
| \psi \rangle=&\sum_{j=1}^{N_1} \beta_{1}^{j}(t) | g_{1}g_{2}...e_{j}...g_{N_1}...g_{N} \rangle | 0 \rangle \nonumber \\
&+\sum_{j=N_1+1}^{N} \beta_{2}^{j}(t) | g_{1}...g_{N_1}g_{N_1+1}...e_{j}...g_{N} \rangle | 0 \rangle \nonumber \\
&+\sum_{\vec{k}}\gamma_k(t) | g_{1}g_{2}...g_{N} \rangle | 1_{\vec{k}} \rangle \nonumber \\
&+\sum_{m,n=1, m\ne n}^{N}\sum_{\vec{k}} \alpha_{k}^{mn}(t) | g_{1}g_{2}...e_n...e_m...g_{N} \rangle | 1_{\vec{k}} \rangle\, ,
\label{phi}
\end{align}
where states in the first two sums correspond to zero number of photons, while in the third sum the photon occupation number is equal to one and all atoms are in the ground state.
The higher order terms in the last row contain more than one excited atom and virtual photons of negative energy \cite{Svidzinsky2008.PRA}. We note here that due to the presence of two transitions in each atom, both virtual photons with the same energy
(corresponding to the same transition in the neighboring atom) and ones with different energies (corresponding to different transitions) can be exchanged between the atoms.  
If the system is initially prepared in an eigenstate, the time evolution of the coefficients $\beta_{1}^{j}(t)$ and $\beta_{2}^{j}(t)$ are given by $\beta_{1}^j(t)=\beta_1^je^{(-\lambda+\frac{i}{2}\phi)t}$, $\beta_{2}^j(t)=\beta_2^je^{(-\lambda-\frac{i}{2}\phi)t}$, with
 $\phi=\omega_1-\omega_2$ being the single-atom Zeeman splitting. Following the notation in Ref.~\cite{Svidzinsky2010.PRA}, the quantities $\lambda$ are the eigenvalues and  $\text{Re}(\lambda)>0$.

One can show (see Supplemental Material \cite{supplmat} for the detailed analytical derivation) that if the eigenstate of the ensemble of $N$  two-level atoms (in the absence of the magnetic field) is given by $\lambda_0=\Gamma+i\mathcal{L}$ \cite{Svidzinsky2010.PRA,Scully2009.superradiance}, where the real part $\Gamma$ stands for the superradiant decay rate and the imaginary part $\mathcal{L}$ represents the collective Lamb shift, then the two eigenvalues in the ensemble with magnetic splitting are given by
\begin{equation}
 \lambda_{\pm}=\frac{\Gamma+\gamma+i\mathcal{L}\mp i\sqrt{\phi^2+[\mathcal{L}-i(\Gamma-\gamma)]^2}}{2}\, .
\label{d27}
\end{equation}
In the following we consider the case of a uniformly excited sample, as could be achieved for instance in thin-film planar cavities in the experiment reporting the observation of collective Lamb shift in single-photon superradiance \cite{rohlsberger2010collective}. This corresponds to assumming that $\beta_1^j(0)+\beta_2^j(0)=\sqrt{\frac{1}{N}}e^{ik_0\cdot \vec{r}_j}$ with $j=1,2....N$. Here  $k_0$ is the wave vector corresponding to the two-level system angular frequency $\omega_0$ in the absence of the magnetic field and $k_0=(\omega_1+\omega_2)/(2c)$, with $c$ denoting the speed of light. Furthermore, we also focus on the limit of a small sample, where the sample size $R$ obeys $k_0R\ll 1$. In this case, we can deduce the field state (see Supplemental Material \cite{supplmat})
\begin{eqnarray}
 | \gamma_0 \rangle&=&\sum_k\sqrt{N}g_ke^{i(k_0-k)\cdot \vec{r}_0}\left(\frac{A_+}{\nu_k-\omega_0+i\lambda_+}\right.\nonumber \\
&&\left.+\frac{A_-}{\nu_k-\omega_0+i\lambda_-}\right) | 1_k \rangle\, ,
\end{eqnarray}
where $A_\pm=\pm(\lambda_\pm-\gamma)/(\lambda_+-\lambda_-)$. For times long compared to the superradiance decay $t\gg\Gamma^{-1}$, we have $| \psi(t) \rangle\rightarrow| g_{1}g_{2}...g_{N} \rangle | \gamma_0 \rangle$ \cite{ScullyZubairy}. The state  $| \gamma_0 \rangle$ is a linear superposition of the single-photon states with different corresponding associated wave vectors, for atoms located all approximately at position $r_0$ in the limit of $k_0R\ll 1$. In this limit, and assuming that 
the coupling constant $\sqrt{N}g_ke^{i(k_0-k)\cdot \vec{r}_0}=g_0$ does not change as long as the frequency of the radiation photon is around the resonant energy, we can define the amplitude of the radiation photon in the mode $k$ as 
\begin{equation}
\sigma_k=\frac{A_+}{\delta-i\lambda_+}+\frac{A_-}{\delta-i\lambda_-}\, ,
\label{d32}
\end{equation}
where $\delta=\omega_0-\nu_k$.  Then the field state $| \gamma_0 \rangle$ can be written as 
$ | \gamma_0 \rangle=-g_0\sum_k\sigma_k| 1_k \rangle$. The amplitude defined in Eq.~(\ref{d32}) can be considered as the atomic response of the system. As a function of the complex variable $\delta$,   $\sigma_k$ has two poles, $\delta_\pm=i\lambda_\pm$.
These poles produce the resonant contributions to the amplitude and hence can be attributed to the effective states with frequencies and dephasing rates given by the real and imaginary parts of $\delta_\pm$, respectively. The amplitude is presented as a superposition of two resonant responses associated with the transitions from the ground state to the corresponding eigenstates. We recall that $\phi$ is the energy difference between the two resonant transitions of the single atom caused by Zeeman effect. Due to  collective effects in the ensemble of $N$ atoms,  the resonant energy difference equals $\sqrt{\phi^2+\mathcal{L}^2}$ when $\Gamma-\gamma=0$ according to Eq.~(\ref{d27}). In the following we extend our analysis for several choices of parameter sets.  

\begin{figure*}[t]
\centering
\includegraphics[width=0.95\textwidth]{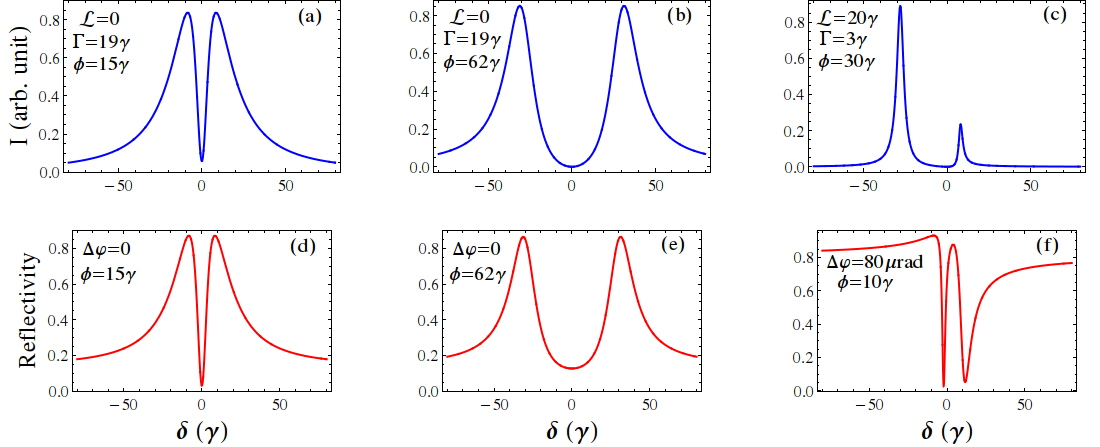}
\caption{\label{fig2} (Color online) (a-c) The radiation spectrum for the three cases discussed in the text: a) EIT-like spectrum for $\Gamma=19\gamma$ and $\phi=15\gamma$, b) regular magnetic splitting with  $\Gamma=19\gamma$ and $\phi=62\gamma$ resembling the single-atom case and c) deviating magnetic splitting for $\Gamma=3\gamma$ and $\phi=30\gamma$. (d-f) Reflectivity of the thin-film cavity for  qualitatively similar cases: d) EIT-like spectrum at resonance incidence angle and $\phi=15\gamma$ corresponding to $B=8$~T, e) regular magnetic line splitting at resonance incidence angle and strong magnetic field $B=33$ T and f) the case of interest with large deviations from the single-atom case at $B=5.3$ T and incidence angle $\Delta\varphi=80$ $\mu$rad.  }
\end{figure*}

We start by considering the particular case when the collective Lamb shift vanishes, $\mathcal{L}=0$. The amplitudes in Eq.~(\ref{d32})  depend on the parameter $x=(\Gamma-\gamma)/\phi$ (we assume $\phi>0$) \cite{anisimov2008decaying}
\begin{equation}
A_\pm=0.5\pm i\frac{x}{2\sqrt{1-x^2}}\, ,
\end{equation}
and the resonance position of the eigenstates in turn on the parameter $y=\sqrt{\phi^2+4\gamma\Gamma}/(\Gamma+\gamma)$ with
\begin{equation}
\delta_\pm=0.5(\Gamma+\gamma)\left(i\pm\sqrt{y^2-1}\right)\, .
\label{d35}
\end{equation}
In the case of significant superradiance, i.e., $\Gamma\gg\gamma$ and $0<y<1$, the real parts $\text{Re}(\delta_\pm)=0$. Hence, both resonances are centered at the resonant transition energy $\omega_0$ in the absence of the magnetic field. The transition line is therefore no longer split by $\phi$ as in the case of the traditional Zeeman effect.  
In turn, the imaginary parts of the poles describe the widths of the eigenstates. In the limit $x\gg1$, the poles according to Eqs.~(\ref{d35}) can be approximately written as
\begin{equation}
\delta_+=i\left(\Gamma-\frac{\phi^2}{4(\Gamma-\gamma)}\right)\, ,
\end{equation}
and 
\begin{equation}
\delta_-=i\left(\gamma+\frac{\phi^2}{4(\Gamma-\gamma)}\right)\, .
\end{equation}
The expressions above show that one of the eigenstates is broad and the other is narrow. In the finite range of $\Gamma\gamma\leq\phi^2/4<(\Gamma-\gamma)^2$ \cite{anisimov2008decaying}, the radiation spectrum which is the sum of broad and the narrow poles results in the characteristic feature of electromagnetically induced transparency (EIT) \cite{EITReview2005}, as illustrated for a numerical example in Fig.~\ref{fig2}a). This spectrum is no longer the sum of the two Lorentzians split by $\phi$ with the effective width $(\Gamma+\gamma)/2$. In fact the non-absorbing feature originates from the difference of two Lorentzians centered at the same position, rather than the summation of two Lorentzians shifted by $\phi$. This clearly reflects the importance of interference. 
In the framework of thin-film x-ray cavities, this interference effect reminding of EIT has been investigated experimentally \cite{rohlsberger2012electromagnetically,slowlight2015} and theoretically \cite{heeg2013x,Heeg2015,Kong2016}.

For significant energy difference $\phi$, in the limit $x\ll1$, the poles can be approximately written as \cite{anisimov2008decaying}
\begin{equation}
 \delta_\pm=\pm\left(\frac{\phi}{2}-\frac{(\Gamma-\gamma)^2}{4\phi}\right)+i\frac{(\Gamma+\gamma)}{2}\, ,
\end{equation}
and the amplitudes are given by
\begin{equation}
 A_\pm=\frac{1}{2}\pm\frac{\Gamma-\gamma}{2\phi}\, .
\end{equation}
As a result, in this case the spectrum can be approximately considered as the summation of two Lorentzians split by $\phi$ with the width $(\Gamma+\gamma)/2$. This is similar to the well-known single-atom Zeeman effect. A numerical example is depicted in Fig.~\ref{fig2}b). A bifurcation point is the special case for which $x=1$, where the spectrum has a pole of  second order:
\begin{equation}
\sigma_k=\frac{\delta-i\gamma}{\left(\delta-i\frac{(\Gamma+\gamma)}{2}\right)^2}\, .
\end{equation}
In this degenerate case, the presentation of the spectrum as a superposition of two Lorentzians is no longer valid. The resulting radiation spectrum is very similar to the EIT one in Fig.~\ref{fig2}a).

We now turn to the case of non-vanishing  collective Lamb shift. For a significant collective Lamb shift $\mathcal{L}\gg(\Gamma-\gamma)$ and $\phi\gg(\Gamma-\gamma)$, we find that
\begin{equation}
\delta_{\pm}=\frac{1}{2}\left[-\mathcal{L}+i\left(\Gamma+\gamma\mp\frac{\mathcal{L}(\Gamma-\gamma)}{\sqrt{\phi^2+\mathcal{L}^2}}\right)\pm\sqrt{\phi^2+\mathcal{L}^2}\right]\, .
\end{equation}
The spectrum can be treated as the summation of two Lorentzians which do not overlap. The Lorentzians are split by $\sqrt{\phi^2+\mathcal{L}^2}$ instead of $\phi$ as in the single-atom Zeeman effect, as illustrated by a numerical example in Fig.~\ref{fig2}c).  In the limit of $\mathcal{L}\gg\phi$, the gap is even closer to $\mathcal{L}$ than to the single-atom splitting $\phi$. Furthermore, also the height of the two Lorenzian curves is modified from the case of a single atom. While the case of a single atom resembles Fig.~\ref{fig2}b), with equally high peaks, collective effects introduce an asymmetry in the peak heights as shown in Fig.~\ref{fig2}c).

To summarize, we find that the collective Zeeman splitting coincides with the single-atom one only for the case of large $\phi$ and $\mathcal{L}=0$. For significant superradiance  $\Gamma\gg\gamma$ and $0<y<1$ or for a large collective Lamb shift value  $\mathcal{L}\gg(\Gamma-\gamma)$ and $\phi\gg(\Gamma-\gamma)$ we obtain qualitatively and quantitatively deviating spectra [see Figs.~\ref{fig2}a) and c)]. Probably the most significant signature of the collective Zeeman splitting could be observed in the case of a large collective Lamb shift value $\mathcal{L}\gg(\Gamma-\gamma)$ and $\phi\gg(\Gamma-\gamma)$. We would like to point out that calculations based on a simplification to the two-level system formalism would yield for the considered case with $g_{\vec{k},1}=g_{\vec{k},2}$ the same value $\mathcal{L}$ for both driven transitions, leading to a vanishing contribution of the collective Lamb shift to the magnetic splitting. Our results above show that this is however not the case. In the context of 
thin-film x-ray cavities, Ref.~\cite{heeg2013x} reaches qualitatively a similar conclusion that spectra are bound to deviate from ``a naive sum of Lorenzians''.

\begin{figure}[h]
\centering
\includegraphics[width=0.5\textwidth]{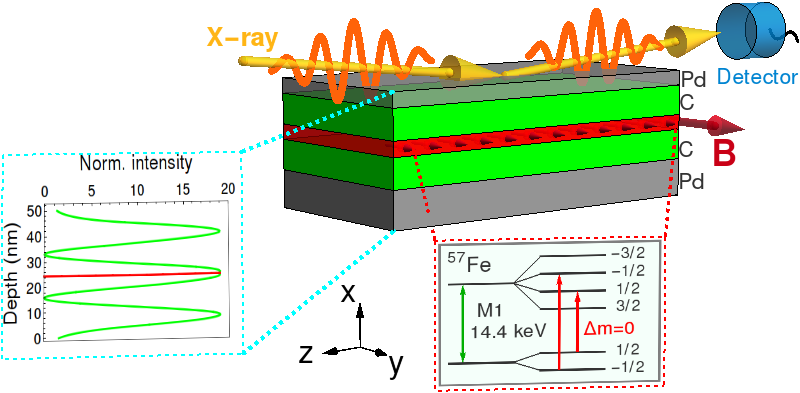}

\caption{\label{fig1} (Color online) Thin-film planar cavity setup with x-ray grazing incidence. The cavity consists of a sandwich of Pd and C layers with a 1 nm $^{57}\mathrm{Fe}$ layer  placed at the antinode of the cavity. The spatial cavity field structure for $\Delta\varphi=0$ is depicted in the inset on the left. The iron nuclei experience a hyperfine magnetic field $\vec{B}$ (red horizontal
arrow) which produces the hyperfine splitting of the levels as illustrated in the inset on the right.}
\end{figure} 

In the following we address the possible experimental observation of the predicted magnetic splitting anomalies in a  thin-film x-ray cavity with a resonant nuclear layer as the one in which the collective Lamb shift was recently observed \cite{rohlsberger2010collective}. Since the resonant system is in this case nuclear and not atomic in nature, we will for convenience refer in the following to {\it magnetic hyperfine splitting} instead of Zeeman splitting.
The planar x-ray cavity consists of  a sandwich of Pd and C layers with a 1 nm $^{57}\mathrm{Fe}$ layer  placed at the antinode of the cavity as illustrated in Fig.~\ref{fig1}. The $^{57}\mathrm{Fe}$ layer is probed by a synchrotron radiation (SR) x-ray pulse at the resonance energy (14.4 keV) at grazing incidence \cite{adams2013}. The evanescently coupled x-ray mode in the cavity satisfies our assumption $k_0R\ll 1$ of a small sample,  the field mode wavelength being much larger than the size of the iron layer, as depicted in Fig.~\ref{fig1}. Typically, the SR pulses contain at most only one photon resonant to the $^{57}\mathrm{Fe}$ nuclear transition. A hyperfine magnetic field perpendicular to the x-ray propagation direction $\hat{k}$  induces the hyperfine splitting of the ground (nuclear spin $I_g=1/2$ and projections $m_g=\pm 1/2$)  and excited ($I_e=3/2$, $m_e=\pm1/2,\pm 3/2$)  $^{57}\mathrm{Fe}$ nuclear states \cite{heeg2013vacuum,Kong2016} as shown in the inset in Fig.~\ref{fig1}. Depending on the polarization of the incident light, specific transitions between the six hyperfine-split nuclear states will be driven. In the following, we consider a linearly polarized x-ray pulse such that only the two $\Delta m= m_e-m_g=0$ transitions  (which have the same coupling constant) can be driven. The nuclear scattering response is measured in the cavity reflectivity signal at the detector \cite{rohlsberger2010collective,heeg2013x,rohlsberger2016strong}.

In the planar cavity system, the collective Lamb shift $\mathcal{L}$ in the absence of the hyperfine splitting is proportional to the ratio $\Delta_C/(\kappa^2+\Delta_C^2)$, where $\Delta_C$ is the detuning of the x-rays from the cavity frequency and $\kappa$ the cavity decay rate. For small angular deviations from the resonance angle $\varphi_0$, the cavity detuning depends on the x-ray photon incidence angle $\varphi$ as  $\Delta_C=-\omega \varphi_0\Delta \varphi$, with $\omega$ the incident radiation frequency and $\Delta\varphi=\varphi-\varphi_0$ \cite{heeg2013x}. Thus, if the incidence angle is exactly on resonance, then the collective Lamb shift equals zero, $\mathcal{L}=0$. 
For off-resonance incidence angles $\varphi\ne\varphi_0$, the Lamb shift has a non-zero value. We use the quantum model developed in Ref.~\cite{heeg2013x} which accounts for the case of a magnetized iron layer to calculate the steady-state cavity reflectivity for three parameter sets which correspond to the generic cases presented in Figs.~\ref{fig2}a)-c). The numerical results for a cavity structure  Pd(5nm)/C(20nm)/Fe(1nm)/C(20nm)/Pd(30nm), incidence angles with $\Delta\varphi=0$ and $\Delta\varphi=80$~$\mu$rad and single-nucleus hyperfine splittings $\phi=10\gamma$, $15\gamma$  and $62\gamma$ are presented in Figs.~\ref{fig2}d)-f). The qualitative agreement between the results of our newly developed formalism and the cavity quantum model \cite{heeg2013x} predictions is very good. The non-resonant reflection channel of the cavity is strongly suppressed in the case of resonant incidence angle $\varphi=\varphi_0$ but present for $\Delta\varphi=80$~$\mu$rad. The interference of the resonant (nuclear) and non-resonant reflectivity channels in the latter case leads to the presence of dips instead of peaks in Fig.~\ref{fig2}f) \cite{heeg2013x,Fano2015}. All quantum model results using the formalism in Ref.~\cite{heeg2013x} have been double-checked and confirmed by comparison with simulations with the software package CONUSS \cite{sturhahn2000conuss} implementing a self-consistent theory including multiple scattering to all orders \cite{Ralf1999}.

We now focus on the results presented in Figs.~\ref{fig2}c) and f) which address the case of a large collective Lamb shift value $\mathcal{L}\gg(\Gamma-\gamma)$ and $\phi\gg(\Gamma-\gamma)$. 
 Since both the superradiant decay $\Gamma$ and the collective Lamb shift $\mathcal{L}$ depend on $\Delta_C$, the two conditions may exclude each other. A close inspection shows that this might be the reason why so far, the collective hyperfine magnetic splitting was never observed in thin-film cavity experiments. For the large magnetic field values used so far in magnetized iron samples, $B\simeq 33$ T, (created by the crystal lattice) the large value of $\phi$ leads to a too small deviation from the single-atom hyperfine splitting. However, for specific off-resonance incidence angles, and lower magnetic field strengths, the conditions $\mathcal{L}\gg(\Gamma-\gamma)$ and $\phi\gg(\Gamma-\gamma)$ can be simultaneously fulfilled. The collective (two-level system) Lamb shift for the incidence angle $\Delta\varphi=80$~$\mu$rad accounts to $\mathcal{L}\approx 6.6\gamma$.  For a magnetic field with $B= 5.3$ T, we find that the anomalous collective hyperfine splitting amounts to  $13.9\gamma$, while the single-atom splitting $\phi=10\gamma$. These values are far enough apart to be tested by an experiment in thin-film x-ray cavities.

In conclusion, we have extended the theoretical formalism for single-photon superradiance for two-level atomic systems to the case of multiplet states introduced by an external magnetic field. 
Our results show that the collective radiation spectrum cannot be accounted for by the simplified picture of independent contributions from several two-level systems. The characteristics of the spectrum can be very useful for quantum optics applications and may also become relevant for studies in M\"ossbauer spectroscopy where the magnetic splitting is used to identify specific compounds.


XK gratefully acknowledges  financial support from the China Scholarship Council.

\bibliography{Zeeman}

\end{document}